\documentclass[journal,compsoc]{IEEEtran}
\usepackage[pdftex]{graphicx}
\usepackage{url}
\usepackage[center]{caption}
\usepackage{todonotes}
\usepackage{pgfplots}
\pgfplotsset{compat=1.13}

\begin{document}

\IEEEoverridecommandlockouts
\IEEEpubid{\makebox[\columnwidth]{\copyright2018 IEEE \hfill} \hspace{\columnsep}\makebox[\columnwidth]{ }}

\presetkeys{todonotes}{inline}{}

\author{Bert Abrath, Bart Coppens, Mohit Mishra, Jens Van den Broeck, Bjorn De Sutter
\thanks{The authors thank the Agency for Innovation by Science and Technology in Flanders (IWT) for supporting Jens and the Fund for Scientific Research - Flanders for project grant 3G013013. Part of this research was conducted in the EU FP7 project ASPIRE, which has received funding from the European Union Seventh Framework Programme (FP7/2007-2013) under grant agreement number 609734.}
}

\title{$\Delta$Breakpad: Diversified Binary Crash Reporting}
\IEEEtitleabstractindextext{%
\begin{abstract}
This paper introduces $\Delta$Breakpad. It extends the Breakpad crash reporting
system to handle software diversity effectively and efficiently by replicating
and patching the debug information of diversified software versions.  Simple
adaptations to existing open source compiler tools are presented that on the one
hand introduce significant amounts of diversification in the code and stack
layout of ARMv7 binaries to mitigate the widespread deployment of code injection
and code reuse attacks, while on the other hand still supporting accurate crash
reporting.  An evaluation on SPEC2006 benchmarks demonstrates that the
corresponding computational, storage, and communication overheads are small.
\end{abstract}

}

\maketitle

\IEEEdisplaynontitleabstractindextext

\section{Introduction and Motivation}
The monoculture in software, in which identical copies of programs are
distributed to all users, has long been blamed for easing the exploitation of
malware~\cite{Forrest97,Cohen93}. As a mitigation, software diversity has been
proposed~\cite{Baudry15,Larsen14,Larsen15}.
The main goal is to prevent that an identified attack vector can
automatically be scaled up to many systems, thus lowering the expected profit of
attacks. 
As software diversification can protect against
many types of attacks, its use is becoming mandated for more and more
systems. Examples include the requirement in many settings to use Address Space
Layout Randomization (ASLR) and MovieLabs' Specification for Enhanced Content
Protection~\cite{movielabs}. The latter mandates software
diversity and so-called copy and title diversity, albeit without prescribing
specific diversification schemes. 

In practice, however, we observe that few, and only very simple
diversification schemes gain traction. With ASLR, for example, only absolute
addresses are randomized, but offsets within executable binaries remain
constant. These limitations open the door to information leak
attacks~\cite{wars}.

\begin{figure*}[t]
\centering
\includegraphics[width=12.5cm]{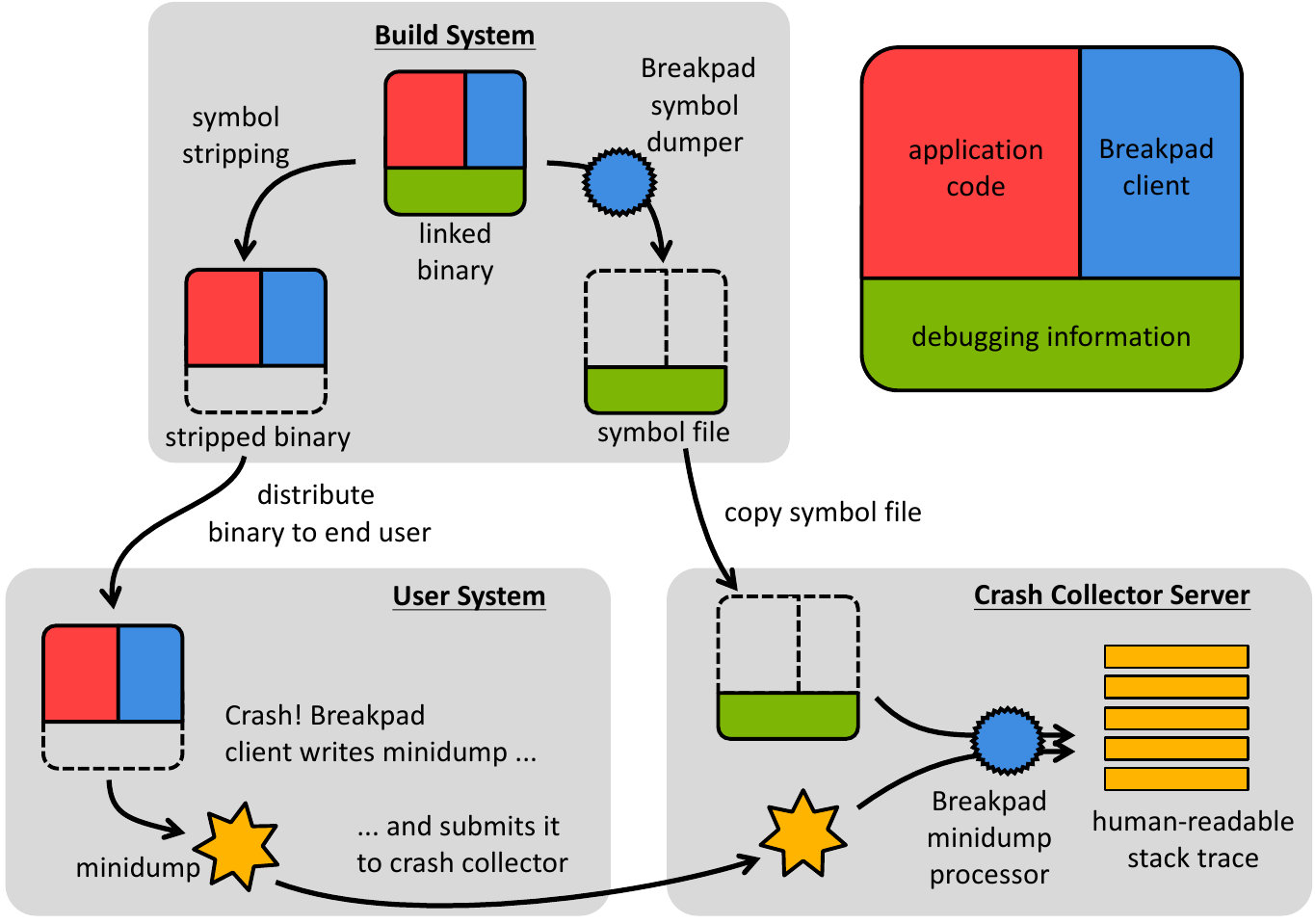}
\caption{Overview of Google's Breakpad tools for crash collection (redrawn after the Breakpad website).}
\label{fig:bp}
\end{figure*}

When academics present new, more advanced diversification schemes, industrial
developers typically appreciate their protection strength, but their costs and
limitations with respect to the software development life cycle (SDLC) severely
restrict their practical usability.
One of the customer support issues relates to crash collectors. Google Breakpad
(\url{http://code.google.com/p/google-breakpad/}), e.g., is a small software
component that can be embedded in applications to facilitate the collection of
crash reports, even when the application binaries are distributed to end
users without debug information. Its operation involving three parties is
visualized in Figure~\ref{fig:bp}. When the application crashes on a user's
system, the embedded Breakpad component sends a stack dump (called minidump) to
the crash collector server. On that server, a tool then combines the minidump
information with the debug information stored in a so-called symbol file on the
server. The tool then generates a stack trace, which most often is first
analyzed and classified automatically. If no equivalent traces are found in a
database of previously received traces, the vendor's developers are notified
that a previously unknown bug or previously unknown trigger has been identified,
at which point they can start to study the trace manually. For obvious reasons,
crash collector tools like Breakpad have become quite popular.

When different users of an application execute different code versions, however,
this system no longer works out of the box. Unless the crash collector stores
symbol files for all of the diversified versions, it lacks the necessary
information to identify and interpret the information in the received
minidumps. According to feedback we get from developers of large, popular open
source projects, simplistic solutions to overcome the mismatch between
diversified minidumps and a single symbol file, such as permanently storing
debug information for all diversified versions, are infeasible because symbol
files are quite large. The alternative solution of rebuilding a software version
and its debug information on the server when a crash report comes in is
considered impractical as well: For larger programs, recompilation of every
crashed version would be compute-intensive, and it requires the precise
reproduction of the developer's build environment in the crash collection
environment, which might reside on a third party's infrastructure. 

Alternatively, we propose to extend the diversified stack dumps with a small
amount of \emph{delta data}~\cite{patent}, which allows the server to overcome
the discussed mismatch without requiring large amounts of persistent
storage, compute power, or communication bandwidth. 

This paper presents such an extension for Breakpad we call $\Delta$Breakpad. It supports crash reporting of binaries diversified with a
combination of three existing diversifications. The contributions of
this paper are the following:
\begin{itemize}
\item An analysis of the effects of three existing diversification schemes on x86
  and ARM debug information.
\item An open-source implementation of those schemes based on minimal adaptations to the widely used, state-of-the-art LLVM 5.0 compiler.  
\item The $\Delta$Breakpad approach, and an open source implementation thereof,
  in which $\Delta$data bridges the gap between a diversified binary crash
  report and debug information from a non-diversified binary. This
  implementation consists of scripts that prepare and manipulate inputs for
  Breakpad components, but it involves no changes to the existing code base.
\item Two techniques to minimize the amount of $\Delta$data necessary
  to bridge that gap.
\item An evaluation on a set of benchmark programs,
  measuring the size of the $\Delta$data, as well as the computational
  cost of building and handling it.
\end{itemize}

The main result is the first demonstration and open source implementation of
co-designed compile-time software diversification on the one hand and crash report server
support for the diversified binaries on the other hand.

This paper is structured as follows. Section~\ref{sec:indirect} provides
background information and analyses the problem to be solved in terms of offset
diversification schemes, debug information required for crash reporting, and the
impact of the diversification on this information, on different types of CPU
architectures. Next, Section~\ref{sec:overview} presents an overview and
detailed discussion of the $\Delta$Breakpad approach as an extension of Google
Breakpad. Section~\ref{sec:setup} discusses practical aspects of the
diversifying tool flow implementation. The results of an experimental evaluation
are presented in Section~\ref{sec:eval}, after which
Section~\ref{sec:discussion} discusses alternative designs and generalization
issues. Section~\ref{sec:related} discusses related work and
Section~\ref{sec:concl} draws conclusions.

\section{Background \& Problem Statement}
\label{sec:indirect}

\subsection{Offset Diversification}
\begin{figure}[t]
\centering
\includegraphics[width=6.7cm]{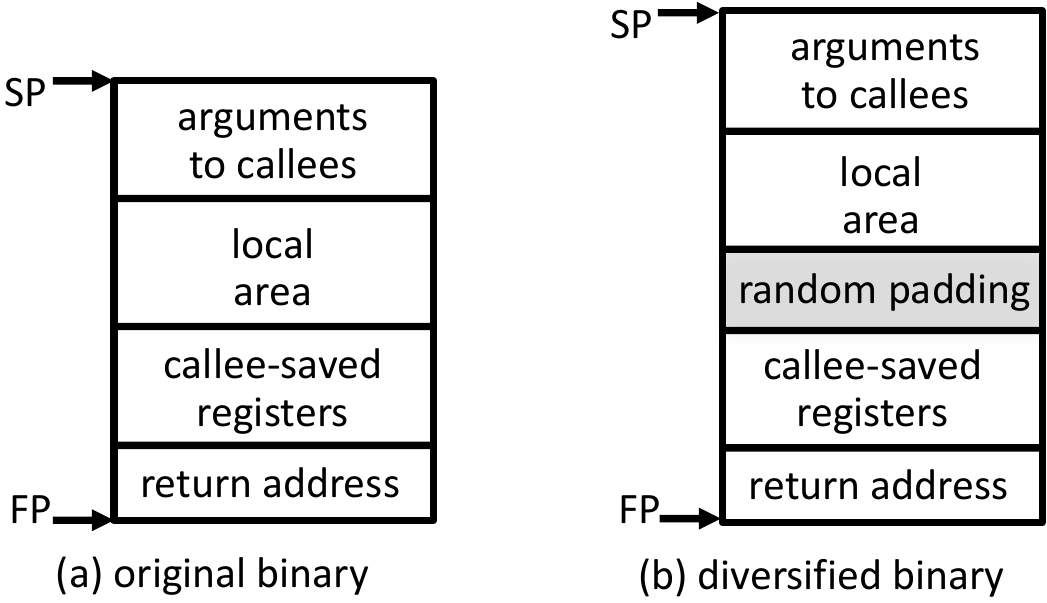}
\caption{Stack frames in original and diversified binaries.}
\label{fig:stack}
\end{figure}

In this work, we focus on diversification schemes that alter offsets between
instructions in a program and offsets between elements in stack frames. We focus
on compiled languages such as C and C++ that provide no memory
safety~\cite{wars}. The studied types of diversification have proven to be
useful on top of basic ASLR, because they raise the bar for information leak
attacks: When offsets within memory segments are diversified on top of their
start addresses, one leaked address no longer directly informs attackers about
the locations of other potentially interesting elements. We deploy
three existing offset diversification schemes:
\begin{enumerate}
\item \textbf{Function Shuffling} The order of all the functions in a whole
  binary is randomized. This randomizes inter-procedural code offsets with high
  entropy~\cite{Kil06}.
\item \textbf{Randomized NOP Insertion} At random locations, for some average
  frequency, NOPs (no-operations) are inserted into the code bodies of all the
  functions. This randomizes intra-procedural code offsets~\cite{Homescu13b}.
\item \textbf{Randomized Stack Padding} A random number of bytes is inserted in
  between the stack locations of buffers and those of the return
  addresses~\cite{Forrest97}. The impact on the stack frames is visualized in
  Figure~\ref{fig:stack}. It randomizes the distance from buffers to stored
  return addresses, as well as the distances between return addresses in
  different stack frames.
\end{enumerate}
We do not claim that these three schemes offer the most powerful protection that
diversification can offer. They do offer significant protection, however, and as
we will demonstrate, can be made compatible with crash reporting.

To implement these schemes, stochastic decision processes decide on the function
ordering, on the locations to insert NOPs, and on the amounts of stack padding
to insert. The stochastic decision processes are deterministic as they are based
on pseudo-random number generators (PRNGs). To generate diversified code
fragments, it suffices to feed the PRNGs different random seeds. 

As the three schemes are conceptually simple, their decision processes do not
involve checks of complex pre-conditions on the code fragments to be
diversified. Hence no complex compiler technology is needed to
replicate the decision processes, even in cases where the application of a
scheme in one compilation step can trigger hard-to-predict indirect effects by
triggering additional code transformations later down the compilation
process. All of the necessary information to replicate them (such as function
names, function body sizes, ...) is readily available in standard debug
information, as will be discussed in the next section, or can trivially be
generated during the compilation process, without needing to make large changes
to the compilers.

A direct effect of the three schemes is that offsets encoded in the code section
of a binary change. With the first two schemes, the displacements between
instructions change, as does the offset of all instructions relative to the
start of the code segment of the binary. In the code section, this implies
changes to the PC-relative offsets encoded in, e.g., direct control flow
transfers. With the third scheme, the direct changes occur in the displacements
between the base pointer and stack pointer on the one hand, and the data items
in a stack frame on the other hand. So offsets encoded in stack memory
operations change, and so do the immediate operands of instructions that produce
pointers to stack-allocated data. In all three schemes, the diversification
hence results in changes to offsets encoded in instructions as immediate
operands. The indirect effect of those changes on the debug information depends
significantly on the type of processor architecture, as we discuss in
Sections~\ref{sec:x86} and~\ref{sec:arm}.

\subsection{Necessary Debug Information}
The debug information of interest is embedded in the symbol
files used by Breakpad. Conceptually, it consists of source line information and
stack unwinding information. For both of those, the code is partitioned in regions: short sequences of consecutive
instructions. The line information consists of a single list of
regions. For each region, the start address, the size, and the corresponding
source file and source line number are stored. In the symbol files that Breakpad
uses, this information is stored in human-readable form, as shown in
Figure~\ref{fig:symbolfile1}. Each line consisting of hexadecimal numbers corresponds
to one region.

\begin{figure}[t]
\small
Description:
{\scriptsize
\begin{verbatim}
FUNC address size parameter_size name 
address size line filenum
\end{verbatim}
}
Example excerpt:
{\scriptsize
\begin{verbatim}
FUNC 157c 34 0 google_breakpad::LineReader::PopLine
157c 4 113 4
1580 30 116 4
FUNC 15b0 38 0 sys_close
15b0 4 2979 16
15b4 1c 2979 16
15d0 10 2979 16
15e0 8 2979 16
FUNC 15e8 5c 0 google_breakpad::PageAllocator::FreeAll
15e8 4 142 13
15ec 8 142 13
\end{verbatim}
}
\caption{Source line mapping in the symbol file.}
\label{fig:symbolfile1}
\end{figure}

The stack unwinding information also consists of a list of regions, described by
their start address and size. Each region also comes with a description of the
locations in the program state where the debugger's stack unwinder will find the
necessary information to unwind the stack. 

Figure~\ref{fig:symbolfile2} shows an excerpt of an ARMv7 symbol file. The post-fix expressions on registers (\texttt{sp}, \texttt{r11},
\texttt{lr}, ...) express how to compute the necessary properties of the frames
on the stack when execution has reached a given region. These properties are the
canonical frame address (\texttt{.cfa}), the return address (\texttt{.ra}), and
the values of callee-saved registers in a function's caller. The first three
records in the excerpt relate to \texttt{function1}, of which the prologue's
assembly code shows it has a frame pointer (FP =\texttt{r11} according to the
ARM EABI). The expression for \texttt{.cfa} on the first line encodes that on
entry to \texttt{function1}, the stack pointer (SP) still points to the start of
the function's stack frame. The second line clarifies that after the push
instruction, two callee-saved registers can be found on the stack, and the SP
points 8 bytes beyond the start of the frame.

\begin{figure}[t]
\small
Description:
{\scriptsize
\begin{verbatim}
STACK CFI INIT address size reg1: expr1 reg2: expr2 ... 
STACK CFI address reg1: expr1 reg2: expr2 ... 
\end{verbatim}
}
Example symbol file excerpts:
{\scriptsize
\begin{verbatim}
STACK CFI INIT 1bdc f0 .cfa: sp 0 + .ra: lr
STACK CFI 1be0 .cfa: sp 8 + .ra: .cfa -4 + ^ r11: .cfa -8 + ^
STACK CFI 1be4 .cfa: r11 4 +

STACK CFI INIT 28a4 f8 .cfa: sp 0 + .ra: lr
STACK CFI 28ac .cfa: sp 20 + .ra: .cfa -4 + ^ r4: .cfa -20 + ^ 
             r5: .cfa -16 + ^ r6: .cfa -12 + ^ r7: .cfa -8 + ^
STACK CFI 28b4 .cfa: sp 904 +
\end{verbatim}
}
Corresponding assembler code excerpts:
{\scriptsize
\begin{verbatim}
<function1>:  push    {fp, lr}
              add     fp, sp, #4
              sub     sp, sp, #16
              ...
<function2>:  push    {r4, r5, r6, r7, lr}
              cmp     r3, #0
              sub     sp, sp, #884    ; 0x374   
              ...
\end{verbatim}
}

\caption{Stack unwinding information in the symbol file.}
\label{fig:symbolfile2}
\end{figure}

To enable the construction of a source-level stack trace on a crash server on
the basis of undiversified debug information and a diversified, crashed binary's
stack dump, $\Delta$Breakpad needs to be able to replicate the diversification's
effect on the symbol file. Given the discussed format of that file,
$\Delta$Breakpad needs to replicate the diversification-induced changes to the
number and ordering of regions, changes to their start addresses and sizes, and
changes to the locations where relevant pieces of program state are stored.

We observed that in the symbol files of our benchmark suites, about 90\% of the
records specify line number information, and about 7\% provide stack unwinding
information, with the rest spend on descriptions of the files and paths, and on
the interfaces that are exported. Those 7\% do occupy about 20\% of the symbol
file size, however: as can be seen in Figures~\ref{fig:symbolfile1}
and~\ref{fig:symbolfile2}, stack unwinding records are much longer than
code/line region records.

\subsection{Indirect effects in x86 binaries}
\label{sec:x86}
On variable-width CISC architectures such as Intel's x86, the indirect effects
of the three diversifications schemes are mostly limited to additional changes
in the displacements between instructions. When, as a result of a changed
offset, less or more bytes are required to encode that offset as an
instruction's immediate operand, the x86 compiler will simply generate another
form of the same instruction that uses less or more bytes. In addition, as the
compiler might put certain instructions on specific alignments to optimize
instruction fetching or instruction caching, it might insert different amounts
of padding as a result of the diversification. Such changes only
alter the addresses and sizes of regions in the symbol files.

More or less the same happens as a result of the randomized stack
padding. In many functions, no instructions are present in the
function prologues/epilogues that only increment/decrement the SP. To
allocated/deallocate the additional randomized padding in such functions,
additional instructions have to be inserted in the prologue/epilogue. In the
symbol file, this comes mostly down to splitting regions in the stack unwinding
information: one region before the SP increment/decrement, and one region after
it.

So replicating the effect of diversification on the debug information stored on
a crash collector requires updating the number, addresses, and sizes of regions,
as well as the offsets where relevant state is stored in stack
frames. To do so, it suffices for the crash collector to have (i) the original,
undiversified binary including its debug information; (ii) a script that replays
the deterministic decision processes of the randomizing diversification schemes;
(iii) the seeds and keys that were used for generating the diversified binary.

So on architectures like the x86, it suffices to
embed the seeds and keys in the binaries, to extend the Breakpad
client to send them along with the minidump to the crash
collector, and to extend the Breakpad minidump processor to let it replicate the
impact of the diversification process on the symbol file. For that replication,
not the whole original compiler is needed. Instead, a simple script suffices
that replays the stochastic diversification decision processes for the program
at hand, i.e., taking into account the alignment requirements of the individual
program fragments and the locations where different types of offsets are encoded
in the code. A complete approach that covers these features and more is
presented in Section~\ref{sec:overview}.

\subsection{Indirect effects in ARMv7 binaries}
\label{sec:arm}
On architectures like the ARMv7 RISC architecture, the situation is quite
different.\footnote{The 32-bit part of the ARMv8 architecture, which is still
  omnipresent on mobile devices, is mostly identical to ARMv7.} The same effect
plays, e.g., with respect to the function prologues and epilogues, but for three reasons there are many more indirect effects. 

\textbf{Fixed-width instruction encoding.} ARMv7 instructions are 16-bit or
32-bit wide. The immediate operands of ALU and LD/ST instructions can therefore
only be quite narrow, so when offsets grow bigger because of diversification, it
can become impossible to encode them as immediate operands. Instead, the offsets
then have to be stored in registers. This requires additional instructions and
puts extra pressure on the register allocator, as a result of which instructions
can also become scheduled in different orders. In fact, we have observed that if
the same offset has to be generated multiple times, the compiler sometimes
applies common-subexpression-elimination~\cite{Muchnick}, which can have a
global impact on register allocation and instruction scheduling. Furthermore, we
have observed that the compiler sometimes changes the base register used in
LD/ST instructions, e.g., when the offsets of a location in the stack frame
relative to the SP and/or the FP change. 

\textbf{Rotating immediate operands.} The ARMv7 architecture has a peculiar way
of encoding offsets as 8 consecutive bits that can be rotated over a 5-bit
amount. It therefore also happens that offsets that could not be encoded as
immediate operands in the original binary become perfectly fine ones after diversification. For example, whereas an original
offset 0x3ff0 cannot be encoded in one immediate operand, it does work perfectly fine for the increased offset 0x4000 that can result from adding stack frame
padding.

\textbf{The visible program counter.} ARMv7 code typically contains a sizable
amount of PC-relative computations, both in position-independent and in
position-dependent code. The reason is the visible program counter
(PC). Constant values that cannot be encoded in individual immediate operands,
such as vectors of numerical values to be used by vector instructions, and
constants unknown at compile time, such as absolute addresses or inter-modular
offsets, are often loaded from so-called literal pools: data chunks dispersed in
between the code that are accessed through PC-relative load operations. As our
diversification schemes can change the sizes of code fragments, and as only
narrow offsets can be encoded, they also affect the location where the compiler
injects the literal pools in between the code. Whereas the order of instructions
and literal pools can remain the same when NOPs are inserted randomly in x86
code, it cannot remain the same in ARMv7 code.

In conclusion, when targeting an architecture like the ARMv7, we have to expect
much further reaching changes to the code section, even if we only
apply our three relatively simple offset diversification schemes. Moreover, on
such an architecture it is impossible to replicate the changes to the
corresponding symbol file completely without replicating part of the compiler
infrastructure that was used during register allocation, instruction selection,
and instruction scheduling. In other words, it cannot suffice to put a simple
script on the crash collector server to replicate the impact of the
diversification on the symbol file.

\section{The $\Delta$Breakpad Approach}
\label{sec:overview}

To overcome this problem, $\Delta$Breakpad combines three
main concepts. The first concept is \emph{imperfect replication} of the
diversification process' impact on the symbol file. 

The second concept is \emph{patching} of the imperfect replication result to make it
perfect. The crash collector will not only receive the necessary seeds and keys
to replicate the diversification decision process, but also a patch that will
allow it to fix any imperfection of the performed replication. So the
$\Delta$Breakpad client has to send both the minidump, the seeds and keys, and
the patch to the crash collector.

The third concept is \emph{$\Delta$-minimization}, with which we denote the
adaptation of the compilation and diversification process to minimize the sizes of the patches
that the client has to send to the crash collector.

\begin{figure*}
\centering
\includegraphics[width=14.5cm]{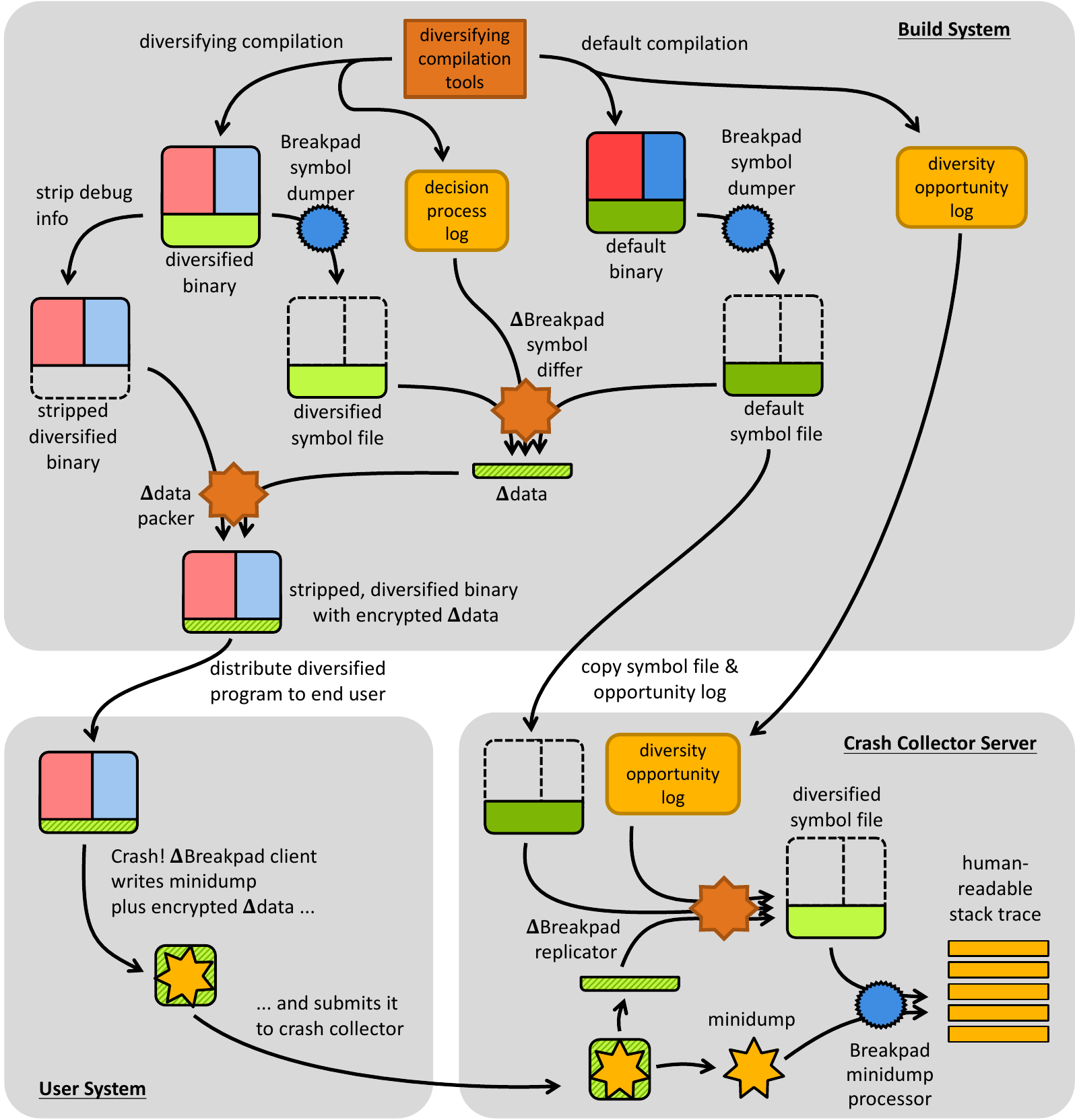}
\caption{Overview of $\Delta$Breakpad as an extension of Google Breakpad. The Breakpad symbol dumper and the Breakpad minidumper are reused as is from the standard Breakpad as shown in Figure~\ref{fig:bp}.}
\label{fig:overview}
\end{figure*}

Figure~\ref{fig:overview} presents an overview of the $\Delta$Breakpad
approach. It looks much more complicated than Breakpad in Figure~\ref{fig:bp},
but the main Breakpad components are still present, and are in fact reused as
is: $\Delta$Breakpad consists of scripts and unmodified existing
Breakpad tools. As we will discuss in Section~\ref{sec:eval}, it requires only
minimal changes to the build system tools to generate the diversified binaries
and $\Delta$data.

\subsection{Crash Handling \& Stack Trace Generation}
\label{sec:crash_handling}

Importantly, the $\Delta$Breakpad approach does not require any change to the
minidump that is sent by the client to the server. The minidump file format as
developed by Microsoft is similar to core dump files, but much smaller, better
documented, and less OS-specific. A minidump contains 
\begin{itemize}
\item A list of the executable and all shared libraries loaded into the process
  when the dump was created.
\item A list of the process threads, with their stacks and processor
  register contents. Complete stacks are included because the applications
  typically do not contain debug information to analyze the stack.
\item Some more system information, incl.\ the processor and OS versions, as
  well as the reason for the crash.
\end{itemize}

We only need adapt the Breakpad client such that it sends the server a small
chunk of \emph{$\Delta$data} along with the minidump (bottom right of
Figure~\ref{fig:overview}). This does not require any patch to the Breakpad
library (\url{https://github.com/google/breakpad/}) that is to be linked into an
application to enable Breakpad crash reporting. That library is only responsible
for dumping the necessary information about a crash to disk. A separate process
is then responsible for sending the data to the crash reporter. This isolation
minimizes the risk that Breakpad's operation is corrupted by the trigger of the
crash (e.g., buggy code being executed). The separate process needs to be
implemented and customized for every OS and usage scenario. For 
$\Delta$Breakpad, we only need to customize it some more to let it
deliver the $\Delta$data with the minidump. That $\Delta$data contains the
random seeds, keys, and other parameters that the server needs to perform the
imperfect replication, as well as the aforementioned patch. If necessary, the
$\Delta$data can be encrypted and signed to guarantee authenticity, integrity,
and confidentiality.

The crash collector server still persistently stores debug info in the form of a
single symbol file of the \emph{default binary}. No changes to its format are
required, so the existing Breakpad symbol dumper utilities for the major OSs can
be reused out of the box to extract the necessary information from the DWARF or
STABS debug sections in ELF object files or from stand-alone PDB (Microsoft's
Program Database format) files.

In addition, the server persistently stores a \emph{diversity opportunity log}.
This log is generated during the \emph{default compilation}, i.e., when the
diversifying tool chain is invoked without applying any actual diversification
to generate the default binary. It lists all the opportunities for
diversification that occurred during the generation of that binary, but that
were not exploited. For example, it lists all the program points where the
diversification process considered but skipped inserting NOPs. An essential
feature of the diversity opportunity log file is that it lists (i) all decision
points where, during an actual diversifying run of the tools, random numbers are
drawn from the PRNG; (ii) the necessary information for determining the
diversification options from which one is selected with each drawn random
number.

When a crash report arrives on the server, the \emph{$\Delta$Breakpad
  replicator} replicates the impact of the diversification process on the symbol
file in a couple of steps. First, the replicator extracts, decompresses, and
(optionally) decrypts the $\Delta$data.

Next, the replicator extracts the seeds, keys and possible parameters from the
$\Delta$data, to replicate the impact of the diversification decision process on
the \emph{default symbol file} by means of the opportunity log. The replicator
initializes a PRNG with the same parameters and random seeds that were already
used on the build system for the actual diversification of the binary from which
the crash report was achieved. The replicator then draws random numbers from
that PRNG at each point where the original diversification process had already
drawn numbers. For each drawn number, the replicator then adapts the content of
the symbol file to replicate (approximately) the impact the original
diversification step had caused on that file. The overall result is an
approximation of the \emph{diversified symbol file}, i.e., the symbol file that
the original Breakpad symbol dumper tool had produced on the build system for
the \emph{diversified binary}. It is an approximation because the replicator
only models direct effects of the diversification, such as increased region
sizes resulting from inserted NOPs, but no secondary effects like the ones
discussed in Section~\ref{sec:arm}. So finally, the replicator extracts the
patch from the $\Delta$data and applies it to the approximation, thus
reproducing an exact copy of the diversified symbol file.

As the contents of that diversified symbol file match the contents of the
received minidump, the existing Breakpad minidump processor can then be used to
produce the human-readable stack trace. Notice that this stack trace only
contains information at the abstraction level of the source code. Crashes
occurring in corresponding regions in differently diversified versions of the
binaries will hence produce exactly the same stack trace. As such, all existing
manual or automatic tools and techniques to analyze and classify the stack
traces, e.g., for triaging, still work out of the box.

\subsection{Generating the $\Delta$data}
The top part of Figure~\ref{fig:overview} shows the adapted build system. On the
right, the standard Breakpad symbol dumper flow is shown to
generate the default symbol file to be stored persistently on the crash collector
server. This symbol file is extracted from the default binary. 


On the left of the build system in Figure~\ref{fig:overview}, the diversified
binary is generated, along with the diversification \emph{decision process log}
that consists of the same info as the opportunity log plus a description of the
actual result from the applied diversification, and a \emph{diversified symbol
  file}. Based on this log and symbol file, and on the default symbol file, our
\emph{$\Delta$Breakpad symbol differ} then generates the $\Delta$data, in
particular the patch part of it. Finally, the \emph{$\Delta$data packer}
compresses, and optionally encrypts and signs the data and injects it as an additional section into
the stripped diversified executable. The resulting binary is then distributed to
the end user, ready to be executed and crash.

\subsection{Combining Multiple Diversification Processes}
\label{sec:combined}
In order to make the described approach work, we need to ensure that the
replication of the decision processes on the crash collector on the basis of the
opportunity log generated for the default binary stays synchronized with the
decision process as it was executed during the generation of the diversified
binary. This is non-trivial when one wants to apply multiple forms of
diversification one after the other. Because of the already discussed indirect
effects of diversifications, the replication process does not know the exact
outcome of an earlier diversification applied to some code fragment. The
replication process hence does not know the exact form of the code fragment onto
which the later diversification is applied.

For example, consider the design where randomized padding is injected into a function's stack
frame first, and random NOPs are inserted in its code body afterwards, after
instruction scheduling has been performed. Given the ordering of compilation
phases in a compiler, this is a reasonable design~\cite{Muchnick}. As
discussed in Section~\ref{sec:arm}, the injected padding can cause changes in
the number of instructions of the function body. If this actually happens, and
if the later NOP insertion process draws a random number for each instruction in
the code to decide whether or not to insert a certain number of NOPs after that
instruction, the replicator will draw more or less random numbers from the PRNG
than were counted during the generation of the default binary. 

Fundamentally, the problem is that the diversifying NOP insertion is then
performed on code that differs from the code from which the opportunity log was
constructed. So in that case, the replication of the decision process on the
crash collector will at some point become desynchronized with how the actual
diversification was decided. Unless special care is taken, this will result in
completely diverging replication from that point on, which can only be
compensated by including a huge patch in the $\Delta$data.

We avoid this in two ways. First, the decision processes of
the combined diversification schemes need to be carefully designed to become
mostly, if not completely independent. In our diversifying tool chain, we
achieve this by applying the later decision processes at a granularity of code
fragments that is not likely impacted by earlier decision processes. Trivially,
the order in which functions are shuffled is completely independent from the
number of NOPs inserted in them, as well as from their stack padding size. We
also observed that although random stack padding and NOP insertion often result
in changes in the number of instructions in the function bodies, in particular
when the ARMv7 architecture is targeted, they rarely impact the structure of the
functions' control flow graphs (CFGs). The few cases in which we did see changes
to the CFGs are the following:
\begin{itemize}
\item When trampolines had to be inserted or could be removed as a result of
  changed code displacements.
\item When basic blocks became so big or small that they (no longer) had to be
  split, e.g., to provide space for a literal pool.
\item When heuristics used by the compiler consider the sizes of
the involved fragments. For example, in the LLVM compiler, we observed that the
\emph{tail duplication} optimization considers code size (small blocks are
duplicated more), as do \emph{if-conversion} and \emph{tail merging}. 
\end{itemize}
Randomized stack padding and NOP insertion can hence impact the CFGs of
functions. Importantly, the effects of the mentioned transformations do not
escape functions, as the transformations are intra-procedural.

Whereas NOP insertion inherently changes the sizes of code fragments, stack
padding changes them much less frequently. We build on this observation by
performing the stack padding insertion first, followed by the NOP insertion, of
which the decision process is performed basic block per basic block, with a
re-initialization of the used PRNG before each block. So however the number of
instructions in the basic blocks are impacted by the former two diversification
steps, as long as the CFG of a function is not impacted, the replicator's
decision process will remain synchronized automatically. Function shuffling is
applied last.

Our second way deals with the above cases where a function's CFG is
actually changed as a result of the first two diversifications. As function
shuffling has no impact on the function bodies, such changes come only from the
stack padding. In such cases, we accept the desynchronization, but we contain it
to the function of which the CFGs are changed, i.e., to that function's part of
the symbol file.

To avoid that the resulting desynchronization in the replication spills over
into other functions, the tools that perform the diversification and the
imperfect replication resynchronize the used PRNGs upon entry to a
function. Such resynchronization per function can be implemented in several
ways. Hierarchical PRNGs are one option, whereby the top-level PRNG is invoked
on entry to each function. In our tools, we alternatively reset the PRNG with a
new seed value that is computed by hashing a unique, immutable identifier of the
function combined with the diversification seeds and keys. With
cryptographically strong hash functions, the new seeds can not be predicted by
attackers unless they know the (global) diversification seeds and key. As a
unique function identifier that is not be impacted by any diversification step,
we use the concatenation of the (mangled) name of the function, the name of the
object file from which the function originated, and the name of its section
within that object file. By compiling code with the \texttt{-ffunction-sections}
flag, these identifiers are guaranteed to be unique. Every function is then put
in its own section in the generated object file, and that section name then
includes the function name, even for functions that are themselves anonymous in
the object file, such as C functions declared \texttt{static}).

\subsection{$\Delta$-Minimization}
\label{sec:delta_minimization}
With this paper, we want to demonstrate that crash reporting for
diversified software is feasible with limited overhead. So we aim for small
$\Delta$data. 

A first option to reduce the size of the $\Delta$data is to
compress it or to use more efficient encodings for the information that needs to
be stored in the $\Delta$data. Compression and coding are not the focus of this
paper, however, so in the remainder of this paper, we will simply rely on
existing compression schemes to compress information encoded in a custom
developed, but likely suboptimal coding scheme.

A second technique is to adapt the processes that perform the compilation and
diversification. Those processes have an impact on the amount of imperfection in
the replication, i.e., on the $\Delta$ between the diversified symbol files and the
symbol files reconstructed through imperfect replication. Those processes can
hence be tweaked to minimize that $\Delta$, which will in turn lead to a
reduction in the amount of patching information needed in the
$\Delta$data. Tweaking the processes is the option we explore in this
section. 

We opt not to achieve a smaller $\Delta$ at all cost, however. Apart from the
restrictions discussed in Section~\ref{sec:combined}, we do not want to impose
strict limitations on the freedom with which to apply the diversification
schemes. For example, when we let a compiler select a randomized amount of stack
padding for some function, we do not want to restrict its selection to values
that preserve the exact instruction schedules in the function body. Besides helping us to
keep the diversification process decision logic (in the compiler as well as in
the replicator) independent of compiler internals, this ensures that the entropy
generated by means of the diversification does not depend more than strictly
necessary on artifacts of the code being diversified. From the perspective of
security, this is obviously an advantage.

Furthermore, we want to limit the changes we need to make to existing compilers
and related tools used for generating and/or diversifying the binaries. 

What remains then to reduce the $\Delta$, is the selection of the default
compilation strategy and a minimal set of adaptations to the compilation tools
to enforce that strategy. For the three forms of offset diversification we
deploy, we identified two tiny but very useful adaptations.

\subsubsection{Adaptation 1: Default Stack Padding}
\label{sec:default_padding}
The first adaptation is that 8 bytes of stack padding are added in every
function in the default, non-diversified binary. During the diversification
process itself, every function gets a randomized number of padding bytes that is
a strictly positive multiple of 8. This adaptation enforces the insertion of
padding operations in all function versions, i.e., default ones and diversified
ones. It therefore limits the number of cases where the code regions of the
function prologues and epilogues as listed in the default symbol file need to be
split to match the regions in the diversified symbol file (as discussed in
Section~\ref{sec:x86}).

The default padding enforces the inclusion of instructions to allocate and
deallocate stack space in the function prologues and epilogues: the single
prologue then contains one \texttt{add sp, sp, \#const} instruction (or multiple
ones, if the size of that stack space, i.e., the \texttt{const} value, cannot be
encoded as a single immediate operand), and each copy of the epilogues contains
one (or more) \texttt{sub sp, sp, \#const} instructions, both in the default
program version and in the diversified versions. Without the default padding,
many functions in the default binary would not contain such SP
incrementing/decrementing instructions. For those functions, the default padding
minimizes the differences between default and diversified code and their
corresponding regions in the symbol files.

For functions that already allocate and deallocate stack space in the default
binary, adding default padding is useful as well. We observed quite some
functions where the local area of a stack frame only holds relatively large
arrays whose sizes are powers of two. In those functions, the aforementioned
\texttt{const} operands are large values of which the least significant bits are
all zeroes. Those values can hence be encoded as single immediate operands in
the ARMv7 and similar architectures. By adding another 8 bytes of padding, a
lower bit becomes set as well. So then the value can no longer be encoded as a
single immediate operand in the default binary, just like it will likely not get
encoded as a single immediate operand in the diversified binaries, where a
randomized, but still relatively small amount of padding is added. The average
difference between the default binary and the diversified binaries, and hence
the average amount of information to be stored in the $\Delta$data, is hence
reduced. For other functions, such as those with small local areas, the added 8
bytes typically don't impact which offsets can be encoded as immediate
operands. The added 8 bytes then do not offer any benefit, but
they also do not hurt in any way.

Minimizing the differences that randomized stack padding introduces between
default and diversified code fragments is particularly important for the
function epilogues; not only to make the corresponding regions in the symbol
files more similar to one another, but also to limit indirect effects on the
generated code. As a result of the default padding, the epilogues in a function
typically have the same size in the default binary and in the diversified
binaries. Maintaining the same size for epilogues throughout the stack frame
diversification is important for $\Delta$-minimization because the size of basic
blocks, which is the form under which epilogues occur in the diversifying
compiler's intermediate code representation, plays a significant role in the
heuristics that steer some compiler optimizations, as discussed in
Section~\ref{sec:combined}. As a result, the insertion of extra instructions in
the epilogues can result in altered CFGs. The introduction of default padding
reduces the occurrence of such alterations. For the interested reader, the
Appendix provides a quantitative analysis of this effect. In any case, reducing
the number of alterations in the CFGs reduces the number of desynchronizations
during the imperfect replication, thus minimizing the required $\Delta$data.

The 8-byte padding in the default binaries has no impact whatsoever on the size
or on the performance of binaries distributed to end users: The default padding
only influences the default symbol files and the $\Delta$data that will be used
to reconstruct the diversified symbol file. With respect to security, there is
only a small impact on distributed software versions. By excluding the
possibility of adding zero bytes of stack padding to a function, keeping only
the values 8, 16, ..., 256, we reduce the entropy in the stack frame layout of
the diversified binaries from $\ln(33)$ to $\ln(32)$.

Note that this 8-byte padding in the default binary can be
implemented trivially in a diversifying compiler that already injects randomized
stack padding: Default stack padding simply comes down to executing the
diversified stack padding code with a non-diversified amount.

With respect to correctness, we note that by making all diversifying padding
multiples of 8 bytes, the padding does not affect the natural alignment of data
in stack frames. Typically, that data needs 8-byte
alignment or less. This is reflected in the application binary interfaces (ABIs)
we know of, and which impose at most 8-byte alignments.  If data in a stack
frame needs stricter alignment, e.g., because vector instructions will operate
on wider data that needs 128-byte or 256-byte alignments, special constructs
need to be used in the code that achieve such alignments independently of the
address at which the stack frame starts. Such constructs include the use of
alloca \texttt{alloca} or the allocation of a bigger array than needed and then
using only an aligned part in that array of which the starting address is
computed at run time. As such constructs function correctly at whatever allowed
stack frame address, i.e., at any 8-byte aligned stack frame address according
to the ABIs, those constructs survive the addition of randomized amounts of
padding that are multiples of 8 bytes.

One can wonder whether the correctness of special programming constructs such as
tail recursion can be affected by stack padding. We conjecture that
this is not the case when the padding is implemented correctly. For
example, we implement the stack padding insertion by simply asking the compiler
to reserve space for more local variables on the stack as if more local
variables were declared in the source code of the functions. The correctness of
the padding then comes down to the correct implementation of the existing stack
frame allocation in the compiler. As that allocation is a crucial aspect of any
compiler, we can rely on its correctness.

\subsubsection{Adaptation 2: SP/FP-relative access optimization}
\label{sec:SP/FP-opt}
The second adaptation consists of disabling a minor optimization in the
(ARM-specific) compiler backend. When a function has a FP, the compiler back-end
can choose to access data in its stack frame via FP-relative LD/ST instructions
or via SP-relative ones. The decision can take into account the offsets of the
data relative to the FP and to the SP. By choosing the option of which the
offset can be encoded in one immediate, rotating operand (as discussed in
Section~\ref{sec:arm}), the code can be optimized.

After disabling that optimization, the compiler alternates less between
FP-relative and SP-relative addressing as a result of randomized padding. The
diversified binaries therefore become more similar to the default binary, which
ultimately results in smaller $\Delta$data. The appendix backs this up with
quantitative data for the interested reader.

This adaptation is trivial to implement: In LLVM, a one-line edit (to a
condition in an if-statement) suffices. However, unlike the default stack
padding, this tweak does potentially impact performance. In the SPEC2006 C and
C++ benchmarks in our benchmark suite compiled with \texttt{-O2}, we observed no
significant average performance impact: the average execution times increased
with the rather small amount of 0.34\%. For individual benchmarks, disabling
the optimization resulted into anything between a 0.86\% speedup and a 2.70\%
slowdown. These effects are likely caused by accident, such as improved or
worsened instruction cache behaviors that accidentally result from small code
changes, i.e., unintentional and beyond the scope and awareness of the
compiler's optimizations~\cite{Mytkowicz2009}. Still, these numbers indicate
that there can be a small effect, that the software developer in certain
performance critical cases may want to trade-off against the potential benefits
in terms of $\Delta$data size. The latter is evaluated in
Section~\ref{sec:eval}.

With respect to security, this adaptation has no impact: The offsets in the
stack frames do not change because of this optimization, and hence the entropy
resulting from the offset randomization is not impacted. With respect to
correctness, this adaptation has no impact either: We only let the compiler skip
the exploitation of an optimization opportunity. In cases where the
transformation implementing the optimization would be mandatory to generate
correct code in the first place, it can of course still be applied as is. We
know of no such cases, however.

\subsection{Profile-Guided Diversification}
Some diversification schemes can benefit from profile information to reduce the
overhead. For example, the performance overhead of NOP-insertion can be reduced
by concentrating NOPs on infrequently executed program points~\cite{Homescu13b}.
$\Delta$Breakpad supports such profile-guided diversification: As long as both
the default compilation and the diversifying compilation runs are served the
same profile information, the decision process logs and the diversity
opportunity log will be consistent with each other, so the $\Delta$Breakpad
replicator will work just fine.

\section{Prototype Diversification Tool Flow}
\label{sec:setup}
As we want to demonstrate that our approach can work with small
$\Delta$data sizes even on architectures that are harder to target, we evaluated
it on the more challenging ARMv7 architecture. In particular, our
prototype tools support the 32-bit subset of the ARMv7-A
architecure (i.e., excluding 16-bit Thumb and Thumb2 code).

Diversification processes can be applied at many stages during the
SDLC~\cite{Larsen14}. In our prototype implementation, the three diversification
schemes are applied when the binaries are built. The schemes are applied in the
already discussed order using existing open-source compiler tools. 

\subsection{Stack Padding}
First, we adapted LLVM 5.0 for randomized stack padding. All functions get a
random stack padding between 8 and 256 bytes, but always a multiple of 8 bytes,
as discussed in Section~\ref{sec:default_padding}. The amount of padding for
each function is determined by hashing the function's (mangled) name. The
diversification seed is the key to the hash function. In this stateless scheme,
the amount of padding in each function is independent of the order in which
functions are compiled. This further eases the replay on the crash server, for
which all the necessary function names are already present in the default
symbol file.

Our LLVM patch to implement the stack padding and related command-line options
is 41 lines of code in total. The stack padding itself is implemented in the
architecture-independent code of the LLVM compiler pass that inserts function
prologues and epilogues. Amongst others, that pass determines the total size of
each function's stack frame, including the space needed to implement calling
conventions. Our patch extends that computation to insert randomized stack
padding.

On top, a two-line patch sufficed to disable the FP/SP-relative stack access
optimization discussed in Section~\ref{sec:SP/FP-opt}. 

\subsection{NOP Insertion}
We further adapted the LLVM 5.0 ARM backend to perform randomized NOP insertion
and to generate an opportunity log, implementing a decision process as discussed
in Section~\ref{sec:combined}. It inserts a NOP in between every consecutive
pair of instructions in a basic block with a user-controlled probability. For
our experiments, we set this probability to 20\%. More complex schemes, that
introduce more entropy in the offsets between individual instructions in
function bodies can easily be envisioned. Introducing many more NOPs will likely
not be acceptable, however, as it obviously inflates the code size. As long as
the more complex schemes have a decision process along the lines of the one
discussed in Section~\ref{sec:combined}, with a fixed number of random numbers
drawn per basic block, we conjecture that the $\Delta$data size will not be
impacted significantly.

To minimize the side effects of the NOP insertion that would lead to inflated
$\Delta$-data, the NOP insertion is done as late as possible in the compiler
backend. The new NOP-insertion compiler pass is invoked after instruction
selection, if-conversion, instruction scheduling, register allocation, peephole
and other assembly-level optimizations, and code layout; and right before the
very last LLVM ARM code generation pass that inserts literal address pools and
the necessary trampolines. As already discussed in Section~\ref{sec:combined},
that last pass can only be executed while all the basic blocks sizes are being
finalized: Trampoline insertion and literal address pool insertion leads to code
size increases, which might necessitate additional insertions, so they are
performed iteratively until a fix-point is reached. From then on, no extra
insertions can be performed (without risking having to undo and redo the
insertion of pools and trampolines).

To replay the NOP insertion on the server, the opportunity log lists the
functions' code and data blocks, as well as their sizes. The data blocks include
blobs of data that the compiler stores in the code section (for various reasons)
as well as the literal address pools. Those blocks are marked as data, such that
the NOP insertion replay knows to skip them, i.e., not to insert NOPs in
them. The code blocks correspond to the basic blocks in the compiler's
intermediate code representation. To enable the inclusion of all the necessary
information, in particular with respect to literal address pools, the
opportunity log is generated at the end of the trampoline and address pool
insertion compiler pass.

Since the number of instructions per basic block can be different in a
diversified binary as a result of stack padding, the data in the opportunity log
allows for relatively accurate, but not perfect replay on the crash server. The
difference is obviously covered by the patch in the $\Delta$data.

Despite our careful design to obtain accurate opportunity logs, we observed that
in some cases, the logs are not completely accurate. When source code contains
inline assembly fragments, the LLVM code generator handles those mostly as
strings, of which it estimates the maximal code sizes to insert trampolines and
literal address pools as necessary. Most often, those estimates are correct. But
sometimes LLVM overestimates their actual size. This results in
desynchronization during the NOP-replication, because the replication then
inserts NOPs in later blocks at incorrect addresses, resulting in incorrect
updates to the supposedly corresponding regions in the symbol file.

Fortunately, this form of desynchronization occurs infrequently. Most user-space
application and library code (except for the standard system libraries) does not
include inline assembly. In our experiments, only the injected Breakpad
components contained inline assembly. For all but the smallest programs, those
components make up only a tiny fraction of the whole binary. Moreover, the
desynchronization ends at the function boundary, when global resynchronization
is performed anyway. So the overall impact on the sizes of the $\Delta$data is
minimal. 

We conjecture it is possible to eliminate this completely by engineering a way
in which incorrect estimates in the opportunity log are patched on the basis of
an inspection of the actual assembler code generated during the default
compilation. This engineering task is left for future work. 

Another source of errors in the NOP insertion replay, and desynchronization, is
the insertion of the NOPs themselves. These can cause the location of the data
pools inside the function to change, or even cause the sizes of these pools to
change. This form of desynchronization happens rather infrequently.

Our LLVM patch to implement the NOP insertion technique and related
command-line options is 148 lines of code in total, 60 lines of which are used
for outputting the opportunity log.

\subsection{Function Shuffling}
We use the standard GNU linker for shuffling functions. In
preparation for this, we use the \texttt{-ffunction-sections} compiler flag to
ensure that the compiler puts each function into a separate code section in the
generated object files. To perform the actual shuffling, we simply generate a
custom linker script that enforces a shuffled order of all the code sections,
and hence of all functions. The order is determined with a pseudo-random number
generator that is seeded with the diversification seed.

This process builds completely on existing linker functionality. No patch to the
linker source code is needed to let it generate the diversified function
orders. For generating the linker script, we extract all the linked-in functions
from the linker map file. All linkers we know can produce such a file, which
basically documents how the original (i.e., default) linker script was executed
on the linked objects.

To replay the shuffling accurately on the crash server, the information
extracted from the linker map file is needed, i.e., the names and sizes of
linked-in functions, as well as their alignment requirements. These can be
obtained from the linker map file and from the object files generated during the
default compilation: the alignment requirements of functions correspond to those
of their corresponding code sections in the object files. Those section
alignment requirements are explicitly encoded in the object files to allow
correct linking. We extract them to include them in the opportunity log. During
the replay, they are useful to predict the amount of padding that needs to be
inserted before each function in the diversified binary, such that that amount
of padding does not need to be included in the $\Delta$data.

\subsection{$\Delta$data}
The uncompressed $\Delta$data our tools generate contain human-readable ASCII
text. With more engineering, smaller patch sizes can likely be obtained, so the
(compressed) $\Delta$data sizes we report in the next section only put an upper
bound on what could be achieved with a more fine-tuned implementation. If
authenticity, integrity and confidentiality are required for the $\Delta$data it
can also be encrypted and signed. This obviously adds some extra data. For
example, when we experimented with GPG (GNU Privacy Guard,
\url{https://www.gnupg.org/}) to encrypt with AES256 and sign using the SHA-1
hash and RSA, we observed that the $\Delta$data grows with 354--356 bytes
(depending on the needed padding).

\section{Experimental Evaluation}
\label{sec:eval}

\subsection{Benchmarks and Correctness}

For evaluating our approach and the correctness of our implementation, we use
the C and C++ programs from the SPEC2006 benchmark suite. We evaluated the
approach on dynamically linked binaries, all of which also include the BreakPad
client next to the actual code. The dynamically linked, position-dependent
binaries were compiled at optimization levels \texttt{-O1}, \texttt{-O2},
\texttt{-Os}, and \texttt{-O3}.  For all four levels, we evaluated two versions:
with and without the \texttt{-fomit-frame-pointer} option. So in total, we
evaluated the benchmarks on eight compilation flag combinations.

For each of those eight combinations, we diversified the benchmarks using 30
tuples of three random seeds, one for each diversification scheme we
implemented. All diversified versions compiled and executed correctly with our
patches and three-step diversification. Hence our diversification implementation
can be considered validated.

To validate the correctness of $\Delta$Breakpad's crash reporting, we checked
and confirm that the diversified symbol files generated with our server-side
replicator on the basis of undiversified symbol files, the opportunity log, and
$\Delta$data are equivalent to symbol files obtained directly with the symbol
dumper from the debug info in the diversified binaries. We also checked and
confirm that correct source-level crash reports are generated based on the
diversified symbol files and mini dumps that we produced by inducing crashes at
randomly selected program locations in the diversified binaries.

\subsection{Overhead}
\label{sec:overhead}

\begin{table*}[t]
\includegraphics[width=\linewidth]{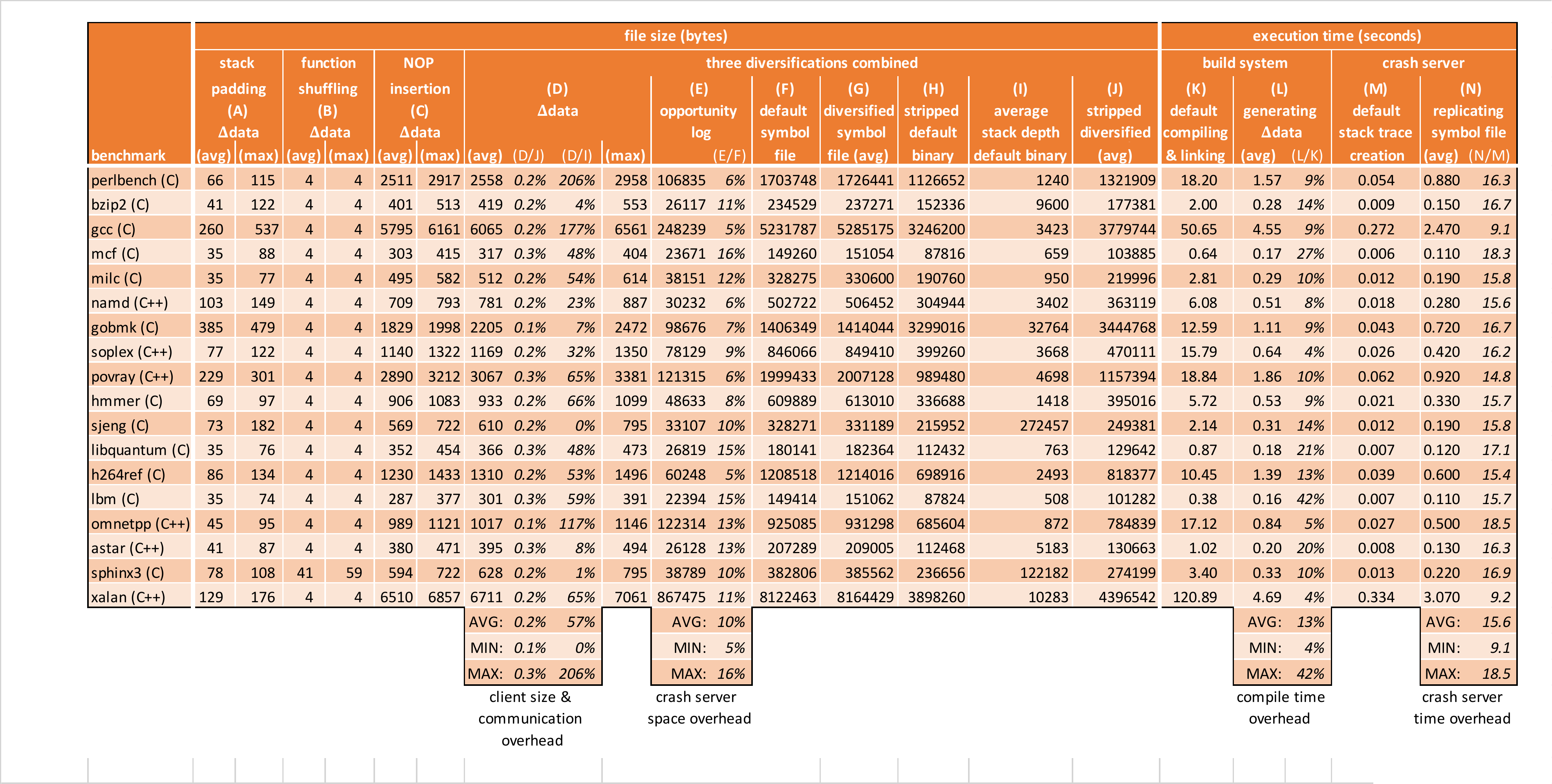}
\caption{Data sizes and execution times for $\Delta$Breakpad use for benchmarks
  compiled at -O2 without FP.}
\label{fig:size_times_results}
\end{table*}

We evaluated the overheads introduced by the diversification and the $\Delta$Breakpad tools with the
two $\Delta$-minimization techniques from Section~\ref{sec:delta_minimization}
enabled. For benchmarks compiled with \texttt{-O2 -fomit-frame-pointer},
Table~\ref{fig:size_times_results} contains the maximum and average sizes of the
$\Delta$data for our three techniques in isolation (A--C) and for all three
combined (D). The listed $\Delta$data sizes are the sizes of the bzipped data,
or simply the size of the random seeds if there was no other $\Delta$data to be
compressed. As the $\Delta$data sizes vary from one diversified version to
another, we list their average size as well as the maximal sizes we observed
during our experiments. These sizes are indicated with ``(avg)'' and ``(max)''
resp. The numbers (E) given for the opportunity logs for three techniques
combined are also compressed using bzip2, as these files are quite large but
very compressible. Also given are the sizes of the default (F) as well as the
diversified symbol files (G), and the sizes of the corresponding stripped
binaries (H and J). For the default binaries, we also report the average stack
depth (I) observed over their execution on SPEC training inputs. This size
corresponds to the amount of stack data that needs to be sent to a crash server
in a minidump. As for the execution times, the table lists the time needed to
compile and link the default binary (K); to generate the $\Delta$data (L); to
create a stack trace for a crash in the main function of the default binary,
which requires no stack unwinding (M); and to produce the diversified symbol
file on the crash server once $\Delta$data is delivered with a minidump
(N). The timing data was gathered using the Python \emph{timeit} module on a
machine with 16 GB of main memory and an Intel i7-4790 CPU. To put the absolute
numbers in the table in perspective, four columns contain relative numbers on
the right and aggregated numbers at the bottom of the table. The formulas to
compute the relative numbers are detailed in the header rows.

We did not include execution times for generating the actual diversification,
because the extra computation time needed to perform the diversification is
negligible compared to the default compilation and linking times.

From the results in Table~\ref{fig:size_times_results}, we can draw several
conclusions. First, the size of the $\Delta$data is small. Even for the three
techniques combined the extra $\Delta$data to be stored in the binaries is
roughly three orders of magnitude smaller than the binary size for each
benchmark. Compared to the average stack size, which is a good indication of the
average size of minidumps to be send to a server, the $\Delta$data can range
from negligible for the sjeng benchmark to relatively large, such as for
perlbench benchmark. Thus, the need to send $\Delta$data can significantly
increase the amount of data to be send to the crash server, up to a factor 3 for perlbench. However, the
increase is relatively high only for programs with shallow stacks. The absolute
increase is, in each case still limited to less than seven kilobytes.

Secondly, the symbol files barely increase as a result of diversification, and the
opportunity logs are about an order of magnitude smaller than the symbol files.
We can thus conclude that on the client as well as on the server, only a
relatively small price is paid in terms of storage for allowing diversified
symbol files to be recreated.

Thirdly, the computation times required to produce the $\Delta$data on the build
system and to produce the diversified symbol files on the crash collector server
are significant. An important remark needs to be made, however. Both the
generation of the $\Delta$data on the build system and the reconstruction of the
diversified symbol file on the crash collector are currently implemented in
Python. Most of the execution time is spent in reading and parsing the default
symbol file, and in allocating the internal data structures that represent
it. These steps can be optimized significantly, by preprocessing the default
symbol file such that it can be mapped into memory with one file open operation,
by re-implementing the scripts in a performance-oriented programming language,
and by redesigning the internal data structures for performance instead of
research flexibility. The reported processing times are therefore only a large
over-approximation of what more fine-tuned implementations will be able to
achieve. We are hence confident that the computational overhead on both the
build system and the crash collector server can be reduced to acceptable
levels. With a reduction with one order of magnitude, which certainly seems
within reach, the overhead on the crash server could be reduced to approximately
a doubling of the computation time needed to produce a crash report.

Fourthly, the observations for C++ programs are in line with those for C
programs. 

Fifthly, from the individual results in columns A--C, we can make several
interesting observations. Stack padding requires significant but relatively
little $\Delta$data. This results from the fact that with the default stack
padding discussed in Section~\ref{sec:default_padding}, relatively few changes
to additional code regions result from stack offset changes. For almost all
benchmarks, function shuffling only requires 4 bytes of $\Delta$data, which are
needed to store the key used for the diversification. For one benchmark,
sphinx3, more $\Delta$data is needed. This results from a small number of system
functions being linked in from pre-compiled crt*.o files, that do not feature
separate sections for each function. As a result, the alignment requirements of
the functions are not replayed correctly, and patching is needed
instead. Finally, the NOP insertion is responsible for the vast bulk of the
$\Delta$data. This is the case because NOP insertion affected the location of
literal address pools in ways that the simple server-side replay cannot predict
accurately.

Figure~\ref{fig:correlation_sizes} charts the main result, i.e., the
$\Delta$data size, in function of the default binary code size for different
compiler optimization levels (always with the $\Delta$-minimization techniques
enabled). The correlation between the two attributes of code size and
$\Delta$data sizes is clear, and it is also clear that the results are quite
similar for the different optimization levels, with or without FP. 

\usetikzlibrary{shapes,backgrounds}

\begin{figure}[t]
\centering
\begin{tikzpicture}
  \pgfplotsset{every axis legend/.append style={
    at={(0,1)},
    anchor=north west}
  }
  \begin{axis}[
      xlabel={Default binary code size (MBytes)},
      ylabel={Average $\Delta$data size (KBytes)},
      legend columns=2,
      legend cell align=left,
    ]
    \addplot[only marks, mark=*, blue] table {results/nofp_-O1/scatter.dat};
    \addlegendentry{-O1, w/o FP\ }
    \addplot[only marks, mark=square*, cyan] table {results/fp_-O1/scatter.dat};
    \addlegendentry{-O1, w FP}
    \addplot[only marks, mark=*, red] table {results/nofp_-O2/scatter.dat};
    \addlegendentry{-O2, w/o FP\ }
    \addplot[only marks, mark=square*, magenta] table {results/fp_-O2/scatter.dat};
    \addlegendentry{-O2, w FP}
    \addplot[only marks, mark=*, yellow] table {results/nofp_-O3/scatter.dat};
    \addlegendentry{-O3, w/o FP\ }
    \addplot[only marks, mark=square*, orange] table {results/fp_-O3/scatter.dat};
    \addlegendentry{-O3, w FP}
    \addplot[only marks, mark=*, green] table {results/nofp_-Os/scatter.dat};
    \addlegendentry{-Os, w/o FP\ \ \ \ }
    \addplot[only marks, mark=square*, lime] table {results/fp_-Os/scatter.dat};
    \addlegendentry{-Os, w FP}

  \end{axis}
\end{tikzpicture}
\caption{Correlation binary code size and $\Delta$data size.}
\label{fig:correlation_sizes}
\end{figure}
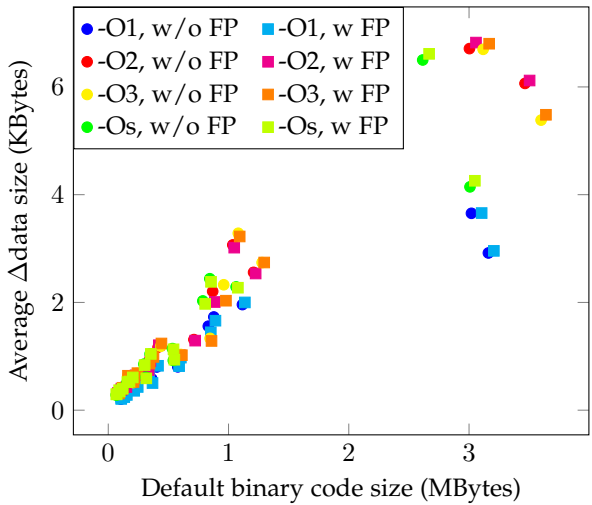

\begin{figure*}[t]
\centering \includegraphics[width=12cm]{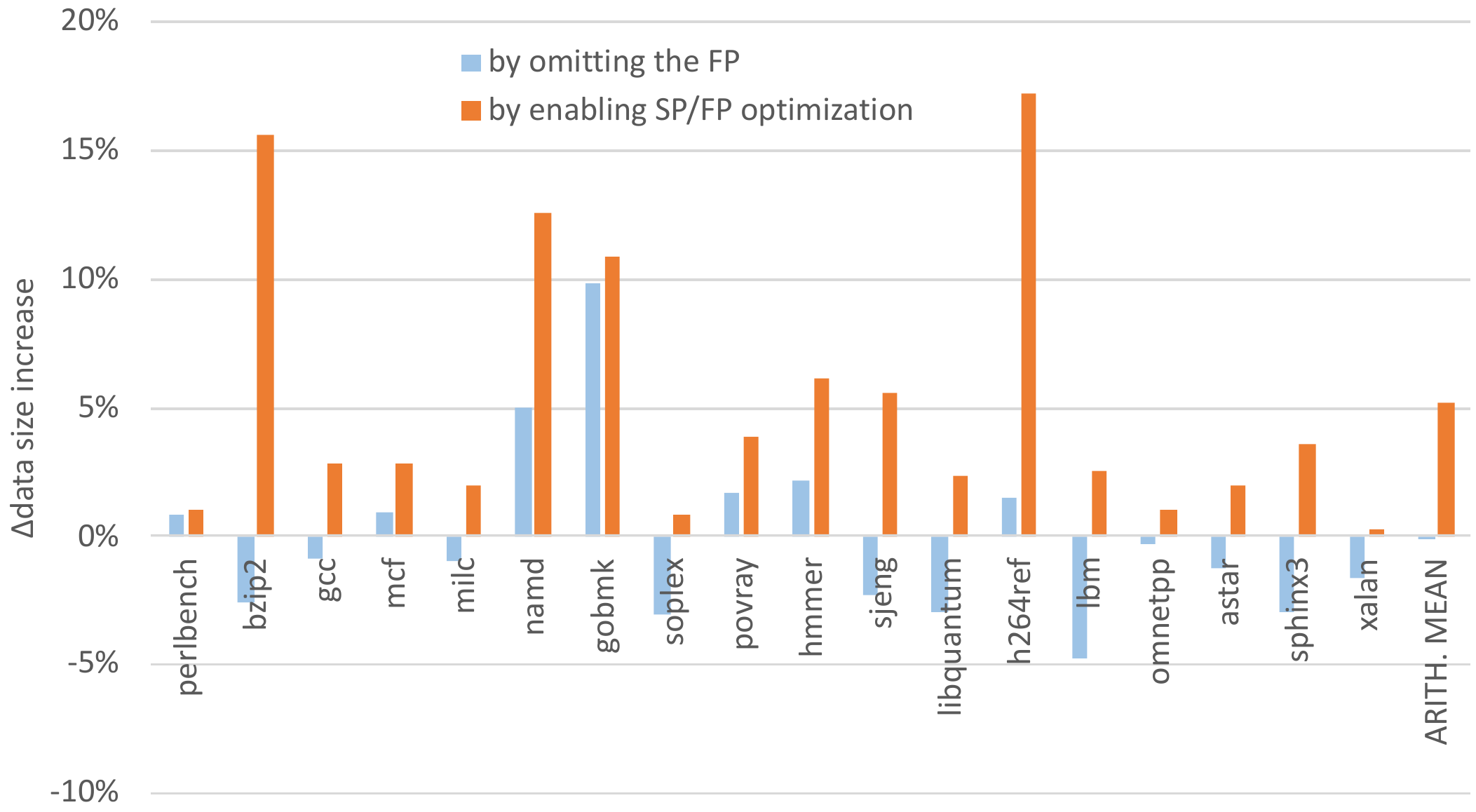}
\caption{Impact on $\Delta$data sizes from omitting the FP and from disabling the FP/SP optimization (on benchmarks compiled with FP) in LLVM (Section~\ref{sec:SP/FP-opt}) for benchmarks compiled at -O2.}
\label{fig:FP_results}
\end{figure*}

Finally, Figure~\ref{fig:FP_results} visualizes the effect on average
$\Delta$data sizes for each benchmark compiled with -O2 ---similar results are
obtained at other optimization levels--- of omitting FPs where possible, and of
deploying the $\Delta$-minimization technique discussed in
Section~\ref{sec:SP/FP-opt}. We did not include the effect of default padding
(Section~\ref{sec:default_padding}) because that does not involve any trade-off,
as it does not effect the diversified binaries themselves. The blue, left bars
indicate the effect on $\Delta$data size of omitting the FP in functions where
that is possible. On some benchmarks, this reduced the size; on others it
increases the size. On average, the effect is negligible. The right, orange bars
indicate the effect on $\Delta$data size of enabling LLVM's SP/FP optimization
when code with FP is generated for all functions. On average, enabling that
optimization leads to 5\% larger $\Delta$data, without outliers up to 18\%. We
conclude that disabling the SP/FP optimization is a useful form of
$\Delta$-minimization for scenarios in which, for whatever reason, developers
insist on letting their compilers generate code with FPs.

Because the whole $\Delta$data of a diversified benchmark version is more or
less equal to a concatenation of $\Delta$data chunks of the benchmark's
functions, and because the effects of ommitting the FP and of disabling the
SP/FP optimization are also local to functions, the absolute effect of those
compilation options on a benchmark's total $\Delta$data size is also mostly a
sum of their effects on a large amount of individual functions. If we assume
that the large set of functions in our benchmark suite is partitioned randomly
into the sets of functions of the individual benchmarks, we expect the results
shown in Figure~\ref{fig:FP_results} to look more like Gaussian distributions
than like uniform ones. And that is what we see. We conclude that if one's goal
is to minimize the $\Delta$data size even further than what we did, the compiler
options should not be enabled or disabled per benchmark. Instead a choice should
be made for each individual function. With machine learning, or maybe even
simple human analysis and engineering, we conjecture that it will be relatively
straightforward to adapt a compiler for this goal. Still, it would be much more
intrusive than the small patch we now deployed to let LLVM inject the randomized
stack padding, the NOP insertion, and the $\Delta$-minimization. So a trade-off
needs to be made. Given the already small sizes of the $\Delta$data with our
implementation, we considered it not interesting to investigate this any further
as of yet.

\section{Discussion}
\label{sec:discussion}

\subsection{Alternative Designs}
In an alternative design option of our approach, one could embed a unique ID in
each diversified binary version, store all $\Delta$data of all program versions
persistently on the crash server instead of in the diversified binaries on the
user systems, and include IDs in delivered crash reports to let the crash server
look-up the corresponding $\Delta$data. The IDs could then also serve as
decryption and signature keys, such that the data on the crash server remains
confidential until it is truly needed to build a crash report.

Despite the small sizes of the required $\Delta$data, one problem of such a
design might be the required storage for all that $\Delta$data. In our design
with the $\Delta$data stored in the binary on the user system, the storage space
occupied by old $\Delta$data is automatically freed as soon as an old binary is
discarded by the user, such as when an application is uninstalled or replaced by
an updated version. No third party needs to be informed when such actions take place. 

If the $\Delta$data is stored on a server instead, the server either needs to
hold on to multiple past and present versions of all $\Delta$data, or it needs
to be informed about the discarding of old binaries by users. In the former
case, more storage space is needed. The latter case, depending on the
application and usage context, involves the collection and communication of
privacy-sensitive and security-sensitive information. Whether either of those
options is feasible, is an open question.  

In any case, a substantial amount of additional storage would be needed on the
crash server. If a crash report service runs on a (small) farm of servers
or in the cloud, it is also an open question as to what the cost might be of
coupling all servers in the service to the necessary storage at sufficient
throughputs and latencies. Whether or not existing storage-computation solutions
might still suffice is unclear; answering this question is out of this
paper's scope.

In our design, where each contacted crash server receives the minidump and the
$\Delta$data over the Internet, only ``centralized'' access to the default
symbol files and opportunity logs is needed. Our experiments indicated that
accessing the opportunity logs on top of the symbol files (that a standard
Breakpad setup needs to access anyway) on average requires only 10\% more data
to be accessed from the ``centralized'' storage. A 10\% increase definitely is
an extra cost, but it is not likely to void the feasibility of existing
storage-computation solutions.

\subsection{General Applicability}
The top level of our $\Delta$Breakpad implementation is architecture-independent
and compiler-independent. Lower-level components are designed to cooperate with
standard Linux binutils tools such as objdump. On top of that the design of the
$\Delta$Breakpad symbol differ, the $\Delta$Breakpad replicator, and the
$\Delta$data format are architecture-independent and compiler-independent.

The implementation of the replicator and the opportunity log format are clearly
architecture-dependent, however, and have been tuned specifically for the
diversification schemes we deploy. Those schemes were also specifically chosen
for the ease with which their effects could be replicated, resulting in small
patches. In principle we can create patches for any diversification scheme, but
there are some trade-offs. Unless replication is at least somewhat correct,
patches will grow to a size where it would be preferable to simply replace them
by the entire diversified symbol file. In other words, our approach then offers
no benefit. Likewise, if the replication becomes too complex or time-consuming
for a certain diversification scheme, the $\Delta$Breakpad approach loses its
appeal.

Consider, for example, the many diversification schemes discussed in the
systematization of knowledge paper by Larsen et al.~\cite{Larsen15}, which we
mark in italics below. We implemented forms of three of these schemes: stack
padding, which is a form of \emph{Stack Layout Randomization}; function
shuffling, which is referred to as \emph{Function Reordering}; and NOP
insertion, which is a form of \emph{Garbage Code Insertion}. We conjecture that
inserting other forms of garbage code will not result in larger $\Delta$data as
long as a similar amount of code is inserted. We furthermore conjecture that other
forms can be supported with smaller $\Delta$data, because only NOPs (having no
side-effects) can be inserted anywhere in code. Other instructions that have
side-effects when executed, can only be inserted where they cannot be reached,
which is definitely in fewer places. As for stack layout randomization, more
heavy-weight schemes (such as those in which the locations of local variables
and spilled data in a stack frame are permuted) will likely require larger
$\Delta$data, because in such schemes the offsets to the SP change. This is not,
or at least rarely, the case in our scheme.

Other existing schemes would result in no changes to the debug information at
all, and thus do not require any replication or patching. This will, e.g., be
the case for some forms of \emph{Register Allocation Randomization} if the
randomization is limited to code-quality-maintaining randomization, i.e., if no
allocations are chosen that lead to longer code schedules. \emph{Instruction
  Reordering} and \emph{Basic Block Reordering} have mostly local effects and we
conjecture that with enough detail in the opportunity logs, which would hence
become longer, these can be replicated sufficiently well. Schemes that have a
larger impact on the control flow graph --- such as \emph{Inlining} and
\emph{Control Flow Flattening} --- would require significantly more detailed
opportunity logs and replication of compiler internals, and therefore most
likely do not fit our approach.

Our current $\Delta$Breakpad diversification schemes are applied at the
compilation and linking stages of the SDLC. Schemes applied during later stages
form no conceptual problem. When the diversification happens after the binary
has been delivered to the user ---as happens with diversification at
installation time, load time, or even at run time--- replication of the
diversification has to be perfect, or a patch has to be created on the user's
system (using $\Delta$Breakpad and the default symbol file). In either case, the
binary delivered to the user has to be accompanied by an opportunity log to
allow for diversification to happen. The diversification by nature can be
replayed without requiring the full build environment, as long as all sources of
randomness used in the diversification process on the user's system are made a
part of the $\Delta$data.

We conjecture that in such cases small opportunity logs and $\Delta$data will
suffice. This conjecture is supported by the fact that currently proposed forms
of diversification applied late in the SDLC are relatively simple and free of
(more global) side effects as the ones we observed in, e.g., LLVM. The reason is
of course that they need to be deployed very quickly to avoid downgrading the
user experience, and hence without heavy-weight compiler technology that can
rewrite code to compensate for side effects.

Finally, we see no reason why our approach would be limited to specific
compilation tool flows. In fact, before we implemented NOP insertion in LLVM, we
already had an implementation in the post-link-time binary-rewriter
Diablo~\cite{ISSPIT05}. So the three schemes were implemented in three separate
tools: the compiler, the linker, and a binary rewriter. While constructing its
intermediate representation of the binary code, Diablo converts literal address
pool entries into instructions. After implementing the NOP insertion, Diablo
then recreates literal address pools. Whereas LLVM creates the pools per
function, Diablo recreates them more globally, in effect combining pools from
multiple functions into single pools. As a result, much fewer such pools end up
in binaries rewritten (and diversified) by Diablo. The number of replay
desynchronizations therefore was also much smaller in those Diablo-diversified
binaries. As a result, the required $\Delta$data for NOP insertion was on
average 2/3 smaller. For some benchmarks, it was even 90\% smaller. We
eventyally decided to switch to LLVM, however, because LLVM is a mature, widely
used tool, which makes the contributions in this paper readily available to
everyone. This required us to adapt the generation of the opportunity log
generation and the replication only slightly.

The source code of $\Delta$Breakpad and all scripts to reproduce the results
presented in this paper are available at
\url{https://github.com/csl-ugent/delta-breakpad}.


\section{Related Work}
\label{sec:related}

In the past, both spatial and temporal software diversity has been proposed as a
solution to a wide range of problems: Instruction set randomization can prevent,
or at least delay, reverse-engineering and
tampering~\cite{Williams09}. Multi-variant execution can be used to detect
malware intrusions~\cite{Volckaert15}. Limited, rather coarse-grained forms of
run-time randomization, such as address space layout randomization (ASLR), are
widely used and significantly raise the bar for memory corruption
attacks~\cite{ASLR}.  In the academic literature, more fine-grained forms of
diversification have been proposed to raise the bar even
further~\cite{Kil06,Giuffrida12}, including for code dynamically generated with
JIT compilers~\cite{Homescu13}. Dynamic temporal diversity has been proposed to
mitigate timing side channel attacks~\cite{Crane15}. Advanced software
fingerprinting schemes can help in identifying the source of illegitimate
software copies~\cite{Collberg07}. Diversification can prevent collusion
attacks to identify software vulnerabilities based on
patches~\cite{Coppens13}. Some software vendors diversify the code of their
applications when major new versions are released, to hide the location of the
new, valuable functionality in the new versions. Obfuscation tools and other
software protection tools inherently rely on diversification to minimize the
learning capabilities of attackers and to achieve
stealthiness~\cite{Collberg09}. Microsoft diversifies the Window's system call
numbering over time to prevent (malicious and beging) software targeting APIs
they do not want to keep backwards compatible~\cite{Windows}.

With the exception of the latter form of diversification, the other forms can
only provide strong protection if code is diversified, i.e., if the
diversification is not limited to changes in the embedded data. 

\section{Conclusions and Future Work}
\label{sec:concl}

In this paper we presented the $\Delta$Breakpad approach to enable crash reporting on diversified software.
We validated this approach for applications on which multiple fine-grained
layout/offset diversifications are deployed. The tool and diversification techniques require only
minimal adaptations to the build tool chain, and only a small price in storage space and communication bandwidth is
paid to support the approach.


Further improvements to our approach can be made with respect to the employed
diversification schemes. Currently these are rather simple, and it is worthwhile
to investigate whether more complex techniques, such as techniques that can be
deployed at install time or at load time, or even at run time, or techniques
that can stop non-control data exploits, can be supported and whether that will
result in a larger overhead in terms of $\Delta$data.


\bibliographystyle{IEEEtran}
\bibliography{paper}


\appendix[Quantitative analysis for $\Delta$-minimization]

\begin{figure*}[t!]
\centering
\small
(a) Benchmarks compiled without FP, with and without default padding, and with FP/SP-optimization disabled\\
\includegraphics[width=0.97\linewidth]{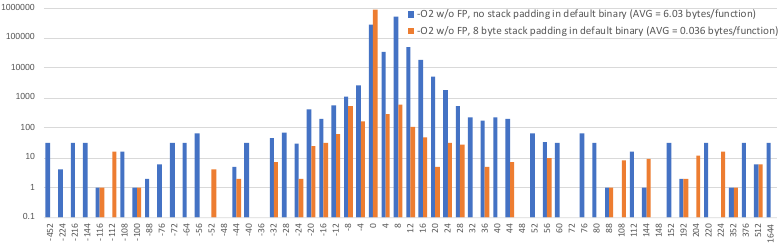}
\vskip 0.2 cm
(b) Benchmarks compiled with FP, with and without default padding, and with FP/SP-optimization disabled\\
\includegraphics[width=0.97\linewidth]{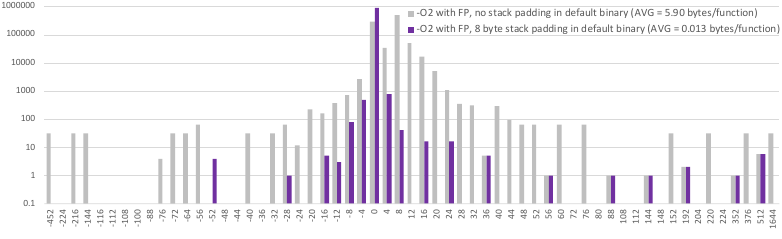}
\vskip 0.2 cm
(c) Benchmarks compiled with FP, with default stack padding, with FP/SP-optimization enabled and disabled\\
\includegraphics[width=0.97\linewidth]{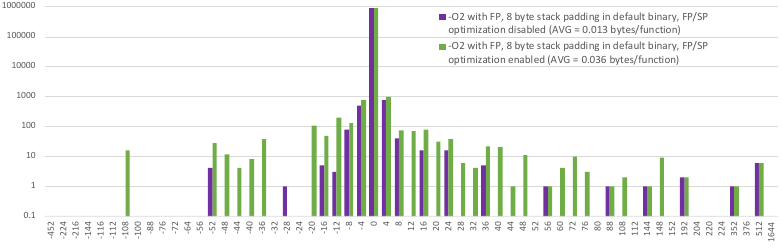}\\
\caption{Histograms of the variation in function size. The Y-axes start at 0.1 to visualize the difference between 0 and 1. The presented average numbers are averages of absolute values of positive and negative variations.}
\label{fig:stack_histogram}
\end{figure*}

Histograms (a) and (b) in Figure~\ref{fig:stack_histogram} quantify the effect
of adding (default) padding to functions on their code size. These histograms
show how the function code sizes change as a result of adding 32 different
amounts of padding (8, 16, ..., 256) to each function in our benchmark suite
compiled with \texttt{-O2 -fomit-frame-pointer} for part (a) and with
\texttt{-O2} for part (b) ---the histograms look similar with other options. The
blue and gray histograms show the changes when the default binary does not
include 8 bytes of padding, the orange and purple histograms show the changes
when the default binary does include 8 bytes of padding.

Notice that many size increases and size reductions are obtained exactly 32 or
64 times in the blue and gray histograms. This follows from the fact that the
same increase or reduction in size was observed for all of the 32
diversified versions of a specific function compared to its default version
without any padding. In the orange and purple histograms, that situation does
not occur. Clearly, the changes on average become much smaller with the default
padding. The average (absolute values of the) changes are 6.03 (respectively, 5.90)
bytes/function without default padding, and only 0.036 (respectively, 0.013)
bytes/function with default padding. Also, the orange and purple histograms peak
at zero, whereas the blue and gray ones peak at 8. So with the default padding,
there are many more functions for which diversified stack padding has no effect
at all on code size. Clearly, the default padding of 8 bytes is advantageous for
$\Delta$ minimization. 

These numbers also indicate that the function size deltas between default and
diversified files are smaller on average for code compiled with FPs than for
code compiled without FPs. The difference is almost completely due to function
versions where the non-zero delta when compiled with FP grows bigger (i.e., more
positive or more negative) in code compiled without FP. The number of function
versions with zero delta compared to the default 8 byte padding version remains
almost constant with or without FP: Over 99.94\% of the 892K function versions
(out of 895k total) that do not grow or shrink in our experiments as a result of
stack padding when compiled with FP, still do not grow or shrink when compiled
without FP.

Histogram (c) in Figure~\ref{fig:stack_histogram} visualizes the effect on
function code size of disabling the SP/FP relative stack access optimization. On
average, the difference in size drops from 0.036 bytes/function to 0.013
bytes/function.

\end{document}